\documentstyle[twocolumn,aps,psfig,epsf]{revtex}

\def\EQ{\begin{equation}}
\def\EN{\end{equation}}
\def\EQA{\begin{eqnarray}}
\def\ENA{\end{eqnarray}}
\begin{document}
\def\eqn#1{Eq.$\,$#1}
\def\mb#1{\setbox0=\hbox{$#1$}\kern-.025em\copy0\kern-\wd0
\kern-0.05em\copy0\kern-\wd0\kern-.025em\raise.0233em\box0}

\title{On the effective velocity created by a point vortex in two-dimensional hydrodynamics}
\author{
Pierre-Henri Chavanis}
\address{Laboratoire de Physique Quantique (UMR5626 du CNRS), Universit\'e
Paul Sabatier, F-31062 Toulouse Cecex 4, France}

\maketitle

\begin{abstract}

We complete previous investigations on the statistics of velocity
fluctuations arising from a random distribution of point vortices in
two-dimensional hydrodynamics. We show that, on a statistical sense,
the velocity created by a point vortex is shielded by cooperative
effects on a distance $\Lambda \sim n^{-1/2}$, the inter-vortex
separation. For $R\gg \Lambda$, the ``effective'' velocity decays as
$R^{-2}$ instead of the ordinary law $R^{-1}$ recovered for $R\ll
\Lambda$. These results are similar to those obtained by Agekyan
[Sov. Astron. 5 (1962) 809] in his investigations on the fluctuations
of the gravitational field.  They give further support to our previous
observation that the statistics of velocity fluctuations are
(marginally) dominated by the contribution of the nearest neighbor.

   \end{abstract}

\vspace{0.1cm}
PACS number(s): 47.10.+g, 47.27.-i, 47.32.-y, 02.50.-r

\vspace{0.4cm}

\narrowtext

Recently, several papers \cite{jimenez}-\cite{kin} have derived the form of the velocity probability distribution function (p.d.f) created by a random distribution of point vortices in two-dimensional hydrodynamics. Technically, the velocity ${\bf V}$ occuring at the center $O$ of a circular domain of radius $R$ is the sum of the velocities ${\mb\Phi}_{i}$ ($i=1,...,N$) produced by the $N$ vortices
\begin{equation}
{\bf V}=\sum_{i=1}^{N}{\mb\Phi}_{i},\qquad {\mb\Phi}_{i}=-{\gamma\over 2\pi}{{\bf r}_{\perp i}\over r_{i}^{2}},
\label{E1}
\end{equation} 
where ${\bf r}_{i}$ denotes the position of the $i$th point vortex relative to the point under consideration and, by definition, ${\bf r}_{\perp i}$ is the vector ${\bf r}_{i}$ rotated by $+\pi/2$ (we have assumed for simplicity that all the vortices have the same circulation $\gamma$). Therefore, if the vortices are randomly distributed over the entire domain, the problem at hand amounts to finding the distribution of a sum of random variables. The difficulty is that the variance of the individual velocities $\langle\Phi^{2}\rangle$ diverges logarithmically so that the Central Limit Theorem is not directly applicable. In early works, Jimenez \cite{jimenez}, Min {\it et al.} \cite{min} and Weiss {\it et al.} \cite{weiss} determined the velocity p.d.f by applying a generalization of the Central Limit Theorem due to Feller \cite{feller} and Ibragimov \& Linnik \cite{ibragimov}. More recently, this problem was reconsidered by Chavanis \& Sire \cite{cs1,cs2} and Chavanis \cite{kin} who extended the methods introduced by Chandrasekhar \& von Neumann \cite{chandra} in their investigations on the fluctuations of the gravitational field. In the stellar context, the limit distribution is a particular L\'evy law, called the Holtzmark distribution. In the vortex context, the velocity p.d.f. is given by 
\begin{equation}
W({\bf  V})= {4\over  {n\gamma^{2}} \ln N} e^{-{4\pi\over n {\gamma^{2}}\ln N}
{V}^{2}}\quad (V\lesssim V_{crit}(N)),
\label{E2}
\end{equation} 
\begin{equation}
W({\bf V})\sim {n\gamma^{2}\over 4 \pi^{2}V^{4}}\quad (V\gtrsim V_{crit}(N)),
\label{E3}
\end{equation}
where 
\begin{equation}
V_{crit}(N)\sim \biggl ({n\gamma^{2}\over 4\pi}\ln N\biggr )^{1/2}\ln^{1/2}(\ln
N).
\label{E4}
\end{equation} 
This distribution is obtained in the ``thermodynamical'' limit $N,R\rightarrow +\infty$ with $n={N\over \pi R^{2}}$ finite (we have assumed, for simplicity, that the vortices are uniformly distributed on average; see Refs. \cite{min,levi,kin} for a generalization of these results to the case of an inhomogeneous medium). In fact, this limit is not well-defined  and the velocity p.d.f. is ``polluted'' by logarithmic corrections. Therefore, in Eqs. (\ref{E2})(\ref{E3})(\ref{E4}) we must consider that $N\rightarrow +\infty$, but not $\ln N$. This is appropriate to physical situations in which the typical number of point vortices does not exceed $10^{4}$. Since this distribution is intermediate between Gaussian and L\'evy laws we proposed to call it a ``marginal Gaussian distribution'' \cite{cs2}. In the (formal) limit $\ln N\rightarrow +\infty$, the tail is rejected to infinity and the limit distribution is Gaussian \cite{feller,ibragimov}. However, the convergence is very slow so that, in physical situations, the algebraic tail is clearly visible \cite{jimenez,min,weiss}. 

From the distribution  (\ref{E2})(\ref{E3})(\ref{E4}) we can easily calculate the average value of the modulus of the velocity. To leading order in $\ln N$, we find
\begin{equation}
\langle |{\bf V}|\rangle=\biggl ({n\gamma^{2}\over 16}\ln N\biggr )^{1/2}.
\label{E5}
\end{equation} 
It should be noted that the moduli of the velocity are {\it not} statistically additive in the sense that
\begin{equation}
\langle |{\bf V}|\rangle\neq \sum_{i=1}^{N}\langle |{\mb\Phi}_{i}|\rangle .
\label{E6}
\end{equation} 
This is clear at first sights since the r.h.s. of Eq. (\ref{E6})
diverges {\it linearly} with the size of the domain.  It is therefore
of interest to calculate the average contribution $\langle \Delta
|{\bf V}|(R)\rangle$ to the modulus of the velocity exerted by
vortices located in the annulus of radius $R$ and thickness $\Delta
R$. A similar quantity was calculated by Agekyan \cite{agekyan} in the
astrophysical context (see also Kandrup \cite{kandrup} for a
generalization of his results to an inhomogenous medium) and it is
possible to extend his method to the case of point vortices. As
we shall see, it is possible to derive an explicit
expression for this quantity in terms of Bessel functions.

Following Agekyan's method, we proceed in three steps: We first determine the average velocity $\langle |{\bf V}^{*}(R)|\rangle$ due to all vortices, except those that lie in the annulus $\Delta R$, assuming that there are $N$ vortices inside $R$. Then, we average this quantity with respect to $N$. Finally, we substract this averaged quantity from the average velocity $\langle |{\bf V}|\rangle$ due to all vortices. Our presentation closely follows the one adopted by Kandrup \cite{kandrup} in his review on the stochastic gravitational fluctuations.  

Suppose there are $N$ vortices inside $R$ and $N_{1}$ between $R+\Delta R$ and $R_{1}$. The probability that these vortices exert a velocity ${\bf V}^{*}$ in $O$ can be obtained by applying Markov's method outlined in Ref. \cite{cs1}. We start from the formula
\begin{equation}
W_{N,N_{1}}({\bf V}^{*})=\int\prod_{i}\tau({\bf r}_{i})d^{2}{\bf r}_{i}\delta\biggl ({\bf V}-\sum_{k}{\mb \Phi}_{k}\biggr ),
\label{E7}
\end{equation} 
where $\tau({\bf r}_{i})d^{2}{\bf r}_{i}$ governs the probability of occurence of the $i$th point vortex at position ${\bf r}_{i}$, and we express the $\delta$ function in terms of its Fourier transform
\begin{equation}
\delta({\bf x})={1\over (2\pi)^{2}}\int e^{-i{\mb\rho}{\bf x}}d^{2}{\mb \rho}.
\label{E8}
\end{equation}
Then,
\begin{equation}
W_{N,N_{1}}({\bf V}^{*})={1\over 4\pi^{2}}\int A_{N,N_{1}}^{*}({\mb\rho})e^{-i{\mb\rho}{\bf V}}d^{2}{\mb\rho},
\label{E9}
\end{equation} 
with
\begin{equation}
A^{*}_{N,N_{1}}({\mb\rho})=\int\prod_{i}\tau({\bf r}_{i})d^{2}{\bf r}_{i}e^{i{\mb\rho}\sum_{k}{\mb\Phi}_{k}}.
\label{E10}
\end{equation} 
Let $i=1,...,N$ label the vortices in the disk of radius $R$ (region (I)) and  $i=N+1,...,N_{1}$ label the vortices between $R+\Delta R$ and $R_{1}$ (region (II)). If the vortices are uniformly distributed in average, then $\tau_{({\rm I})}=1/\pi R^{2}$ and $\tau_{({\rm II})}=1/\pi (R_{1}^{2}-(R+\Delta R)^{2})$. We can write therefore,  
\begin{eqnarray}
 A_{N,N_{1}}^{*}({\mb\rho})=\biggl (\int_{|{\bf r}|=0}^{R}\tau_{({\rm I})}e^{i{\mb\rho}{\mb\Phi}}d^{2}{\bf r}\biggr )^{N}\times
\qquad\qquad\qquad\nonumber\\
\biggl (1-\int_{|{\bf r}|=R+\Delta R}^{R_{1}}\tau_{({\rm II})}(1-e^{i{\mb\rho}{\mb\Phi}})d^{2}{\bf r}\biggr )^{N_{1}},
\label{E11}
\end{eqnarray} 
where
\begin{equation}
{\mb\Phi}=-{\gamma\over 2\pi}{{\bf r}_{\perp}\over r^{2}}.
\label{E11new}
\end{equation}

In the limit $R_{1},N_{1}\rightarrow \infty$ with $n=N_{1}/\pi R_{1}^{2}$ fixed (and treating $\ln N_{1}$ as a constant), we get 
\begin{equation}
W_{N}({\bf V}^{*})={1\over 4\pi^{2}}\int A_{N}^{*}({\mb\rho})e^{-i{\mb\rho}{\bf V}}d^{2}{\mb\rho},
\label{E12}
\end{equation} 
with
\begin{eqnarray}
 A_{N}^{*}({\mb\rho})=\biggl (\int_{|{\bf r}|=0}^{R}{1\over \pi R^{2}}e^{i{\mb\rho}{\mb\Phi}}d^{2}{\bf r}\biggr )^{N}\times
\qquad\qquad\qquad\nonumber\\
{\rm exp}\biggl\lbrace -n\int_{|{\bf r}|=R+\Delta R}^{R_{1}}(1-e^{i{\mb\rho}{\mb\Phi}})d^{2}{\bf r}\biggr\rbrace.
\label{E13}
\end{eqnarray} 
Let $P(N)$ denote the probability of finding $N$ vortices in the disk of radius $R$. Since the vortices are randomly distributed with a uniform distribution in average, $P(N)$ is given by the  Poisson distribution
\begin{equation}
P(N)=(n\pi R^{2})^{N}{e^{-n\pi R^{2}}\over N!},
\label{E14}
\end{equation} 
where $n$ is the average density  of vortices. Averaging the quantity (\ref{E13}) with respect to the Poisson distribution (\ref{E14}), we obtain
\begin{eqnarray}
 A^{*}({\mb\rho})=\sum_{N} P(N)A_{N}^{*}({\mb\rho})\qquad\qquad\qquad\qquad\qquad\qquad\nonumber\\
={\rm exp}\biggl\lbrace -n\int_{|{\bf r}|=R+\Delta R}^{R_{1}}(1-e^{i{\mb\rho}{\mb\Phi}})d^{2}{\bf r}\biggr\rbrace\qquad\qquad\qquad \nonumber\\
\times e^{-n\pi R^{2}} \sum_{N=0}^{+\infty}{1\over N!}\biggl (\int_{|{\bf r}|=0}^{R} n e^{i{\mb\rho}{\mb\Phi}}d^{2}{\bf r}\biggr )^{N}\qquad\qquad\qquad\nonumber\\
=e^{-nC({\rho})}{\rm exp}\biggl \lbrace n \int_{|{\bf r}|=R}^{R+\Delta R}(1-e^{i{\mb\rho}{\mb\Phi}})d^{2}{\bf r}\biggr\rbrace,\qquad\qquad
\label{E15}
\end{eqnarray} 
where 
\begin{eqnarray}
C({\rho})=\int_{|{\bf r}|=0}^{R_{1}}(1-e^{i{\mb\rho}{\mb\Phi}})d^{2}{\bf r}.
\label{E16}
\end{eqnarray} 
The function $C(\rho)$ has been calculated in Ref. \cite{cs1}. To simplify the integral appearing in Eq. (\ref{E15}), we find it convenient to take ${\mb\Phi}$ as a variable of integration instead of ${\bf r}$ (see Ref. \cite{cs1} for more details). Transforming to polar coordinates, carrying out the angular integration and taking the limit $\Delta R\rightarrow 0$, we obtain 
\begin{eqnarray}
 A^{*}({\mb\rho})=e^{-nC({\rho})}\biggl\lbrace 1+2\pi n R\Delta R\biggl \lbrack 1-J_{0}\biggr ( {\gamma\rho\over 2\pi R}\biggr )\biggr \rbrack\biggr\rbrace,
\label{E17}
\end{eqnarray} 
where $J_{0}$ is the Bessel function of zeroth order. Eq. (\ref{E12}), without the subscript $N$, and Eq. (\ref{E17}) determine the distribution of the velocity fluctuations occuring in $O$ due to all the vortices except those located between $R$ and $R+\Delta R$. If we denote by
\begin{eqnarray}
\langle \Delta |{\bf V}(R)|\rangle =\langle |{\bf V}|\rangle-\langle|{\bf V}^{*}|\rangle, 
\label{E18}
\end{eqnarray} 
the average velocity due to the vortices located in the annulus of radius $R$ and thickness $\Delta R$, we have
\begin{eqnarray}
\langle \Delta |{\bf V}(R)|\rangle=-{1\over 4\pi^{2}}\int e^{-i{\mb\rho}{\bf V}}e^{-nC({\rho})}2\pi n R\Delta R\nonumber\\
\times \biggl \lbrack 1-J_{0}\biggr ({\gamma\rho\over 2\pi R}\biggr) \biggr \rbrack Vd^{2}{\bf V}d^{2}{\mb\rho}.
\label{E19}
\end{eqnarray} 
Transforming to polar coordinates and carrying out the angular integrations, we can rewrite the foregoing expression in the form
\begin{eqnarray}
\langle \Delta |{\bf V}(R)|\rangle=-2\pi n R\Delta R \int_{0}^{+\infty}VdV\nonumber\\
\times\int_{0}^{+\infty}\rho V J_{0}(\rho V)g(\rho)d\rho,
\label{E20}
\end{eqnarray} 
with
\begin{eqnarray}
g(\rho)=\biggl\lbrack 1-J_{0}\biggl ({\gamma\rho\over 2\pi R}\biggr)\biggr\rbrack e^{-nC(\rho)}.
\label{E21}
\end{eqnarray} 
Under this form, it is not possible to interchange the order of integrations. However, using the identity
\begin{eqnarray}
xJ_{0}(x)={d\over dx}(xJ_{1}(x)),
\label{E22}
\end{eqnarray} 
and integrating by parts, we obtain
\begin{eqnarray}
\langle \Delta |{\bf V}(R)|\rangle=2\pi n R\Delta R \int_{0}^{+\infty}dV
\int_{0}^{+\infty}  V J_{1}(\rho V)h(\rho)d\rho,
\label{E23}
\end{eqnarray} 
with
\begin{eqnarray}
h(\rho)=\rho g'(\rho).
\label{E24}
\end{eqnarray} 
Integrating by parts one more time with the identity 
\begin{eqnarray}
J_{0}'(x)=-J_{1}(x),
\label{E25}
\end{eqnarray} 
we find
\begin{eqnarray}
\langle \Delta |{\bf V}(R)|\rangle=2\pi n R\Delta R \int_{0}^{+\infty}dV\int_{0}^{+\infty} J_{0}(\rho V)h'(\rho)d\rho.
\label{E26}
\end{eqnarray} 
It is now possible to interchange the order of integrations. Using the identity
\begin{eqnarray}
\int_{0}^{+\infty}J_{0}(x)dx=1,
\label{E27}
\end{eqnarray} 
we get
\begin{eqnarray}
\langle \Delta |{\bf V}(R)|\rangle=2\pi n R\Delta R \int_{0}^{+\infty} {h'(\rho)\over\rho} d\rho.
\label{E28}
\end{eqnarray} 
Straightforward integrations by part lead to the equivalent expression
\begin{eqnarray}
\langle \Delta |{\bf V}(R)|\rangle
=2\pi n R\Delta R \int_{0}^{+\infty} {g(\rho)\over\rho^{2}} d\rho.
\label{E29}
\end{eqnarray} 
Therefore, the average value of the velocity created by the point vortices located in the annulus between $R$ and $R+\Delta R$ is given by 
\begin{eqnarray}
\langle \Delta |{\bf V}(R)|\rangle=2\pi n R\Delta R \qquad\qquad\qquad\qquad\qquad\qquad\nonumber\\
\times\int_{0}^{+\infty}\biggl\lbrack 1-
J_{0}\biggl ({\gamma\rho\over 2\pi R}\biggr )\biggr\rbrack e^{-nC(\rho)}{d\rho\over \rho^{2}}.
\label{E30}
\end{eqnarray} 
The exact expression for $C(\rho)$, calculated in Ref. \cite{cs1}, is relatively complicated. However, for our purposes, it is sufficient to consider its approximate expression
\begin{eqnarray}
C(\rho)={\gamma^{2}\over 16\pi}\ln N_{1}\rho^{2}.
\label{E31}
\end{eqnarray} 
With the change of variables $x=\gamma\rho/2\pi R$, we can rewrite
Eq. (\ref{E30}) in the form
\begin{eqnarray}
\langle \Delta |{\bf V}(R)|\rangle= n\gamma \Delta R \int_{0}^{+\infty} (1-J_{0}(x))e^{-{n\over 4}\pi R^{2}\ln N_{1} x^{2}}{dx\over x^{2}}.
\label{E32}
\end{eqnarray} 
It turns out that this integral can be expressed in terms of modified Bessel functions. Indeed,
\begin{eqnarray}
\langle \Delta |{\bf V}(R)|\rangle= n\gamma \Delta R \biggl\lbrace -\sqrt{\pi a}+{1\over 4}\biggl ({\pi\over a}\biggr )^{1/2}e^{-{1\over 8a}}\nonumber\\
\times\biggl\lbrack (1+4a)I_{0}\biggl ({1\over 8a}\biggr )+I_{1}\biggl ({1\over 8a}\biggr )\biggr\rbrack\biggr\rbrace,
\label{E33}
\end{eqnarray} 
with 
\begin{eqnarray}
a\equiv {n\over 4}\pi R^{2}\ln N_{1}.
\label{E34}
\end{eqnarray} 
For $R\rightarrow 0$, this expression reduces to 
\begin{eqnarray}
\langle \Delta |{\bf V}(R)|\rangle= n\gamma \Delta R.
\label{E35}
\end{eqnarray} 
This formula would be expected to be general if the moduli of the velocity were additive. Indeed, in that case
\begin{eqnarray}
\langle \Delta |{\bf V}|\rangle_{2}= n 2\pi R\Delta R\times {\gamma\over 2\pi R}=n\gamma \Delta R.
\label{E36}
\end{eqnarray}  
However, the moduli of the velocity are {\it not} additive and Eq. (\ref{E35}) does not hold for large values of $R$. Indeed, for $R\rightarrow +\infty$,
\begin{eqnarray}
\langle \Delta |{\bf V}(R)|\rangle={\gamma\over 4}\biggl ({n\over\ln N_{1}}\biggr )^{1/2}{\Delta R\over R}.
\label{E37}
\end{eqnarray} 
It is seen that ``cooperative'' effects modify the expression of the average velocity that we would expect on the basis of naive arguments. In view of the asymptotic behaviors (\ref{E35}) and (\ref{E37}), a simple approximate expression for the average velocity is given by  
\begin{eqnarray}
\langle \Delta |{\bf V}(R)|\rangle={\langle \Delta |{\bf V}|\rangle_{2}\over 1+R/\Lambda}, 
\label{E38}
\end{eqnarray} 
where $\Lambda$ is a characteristic length defined by
\begin{eqnarray}
\Lambda={1\over (16 n\ln N_{1})^{1/2}}.
\label{E39}
\end{eqnarray}

\begin{figure}
%      \vspace{5cm}
   \epsfxsize=8.8cm\epsfbox{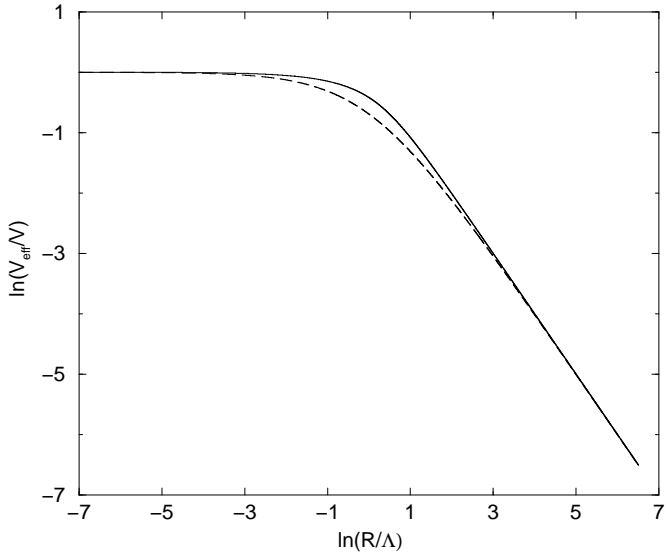} \caption[]{Plot of the effective velocity normalized by the physical velocity $V=\gamma/2\pi R$ versus the normalized distance $R/\Lambda$. The ``effective'' velocity is defined by $V_{eff}=\langle \Delta |{\bf V}(R)|\rangle/2\pi n R\Delta R$.  The full curve corresponds to the exact formula deduced from Eq. (\ref{E33}) and the dashed line to the approximate formula (\ref{E40}). }
           \label{agekyan}
     \end{figure}

This formula shows that the velocity is ``shielded'' on a typical distance $\sim \Lambda$, the inter-vortex distance. Therefore, everything happens {\it as if} the velocity were additive but that each vortex produced an ``effective'' velocity  
\begin{eqnarray}
V_{eff}={\gamma\over 2\pi R}{1\over 1+R/\Lambda}.
\label{E40}
\end{eqnarray}
For $R\gg\Lambda$, the ``effective'' velocity decays like $R^{-2}$ instead of the usual behavior $ R^{-1}$ recovered for $R\ll\Lambda$ (see Fig. \ref{agekyan}).

These results confirms our previous observation \cite{cs1,kin} that the velocity fluctuations occuring in $O$ are dominated by the contribution of the nearest neighbor. In  fact, this dominance is marginal since the average velocity diverges logarithmically with the size of the domain
\begin{eqnarray}
\langle |{\bf V}|\rangle=\int_{0}^{+\infty}{n\gamma\over 1+R/\Lambda}dR\sim \gamma n^{1/2} (\ln R_{1})^{1/2}.
\label{E41}
\end{eqnarray} 
Therefore, equal logarithmic intervals contribute equaly to the velocity. In other words, the contribution from vortices between $R$ and ${R\over 2}$, ${R\over 2}$ and ${R\over 4}$ and so forth are of equal importance. Therefore, the contribution of the nearest neighbor (at a typical distance $\sim n^{-1/2}$ from the point under consideration) $V_{n.n}\sim \gamma n^{1/2}$ is of the same order, up to a logarithmic factor, as the contribution of the rest of the system. {\it In a sense}, we can say that the velocity (\ref{E41}) is dominated by the contribution of the nearest neighbor and that collective effects are responsible for logarithmic corrections. As described in Ref. \cite{cs1}, this particular circumstance arises because we are just at the separation between Gaussian and L\'evy laws.  

The results obtained in this paper and in Refs. \cite{jimenez,min,weiss,chukbar,kuvshinov,cs1,cs2,levi} can have applications in the context of decaying two-dimensional turbulence when the flow becomes dominated by a large number of coherent vortices. They have been used explicitly by Sire \& Chavanis \cite{sc} to build up an effective two- and three-body dynamics of vortices subjected to the ``noise'' induced by the rest of the system. They are also of great importance to build up a rational kinetic theory of point vortices in two-dimensional hydrodynamics \cite{kin}. On a formal point of view, they complete the analogy between 2D vortices and stellar systems investigated in Refs. \cite{these,floride}. They can also be relevant for other fields of physics such as non neutral plasmas under a strong magnetic field \cite{levi} and parallel straight dislocation systems \cite{groma}.

\vskip1cm 
I acknowledge interesting discussion with C. Sire on this topics.

\end{document}